\documentclass[aps,prb,showpacs,twocolumn]{revtex4-1}

\usepackage{amsmath,graphics}
\usepackage[next]{inputenc}
\usepackage[dvips]{epsfig}

\def\be{\begin{equation}}
\def\ee{\end{equation}}

\def\bi{\begin{itemize}}
\def\ei{\end{itemize}}
\def\bn{\begin{enumerate}}
\def\en{\end{enumerate}}
\def\bea{\begin{eqnarray}}
\def\eea{\end{eqnarray}}

\def\ba{\begin{array}}
\def\ea{\end{array}}
\def\bd{\begin{displaymath}}
\def\ed{\end{displaymath}}

\begin{document}
\title{Topological phases and phase transitions on the square-octagon lattice}
\author{Mehdi Kargarian}
\email[]{kargarian@physics.utexas.edu}
\affiliation{Department of Physics, The University of Texas at Austin, Austin, TX 78712, USA}
\author{Gregory A. Fiete}
\affiliation{Department of Physics, The University of Texas at Austin, Austin, TX 78712, USA}

\begin{abstract}
We theoretically investigate a tight binding model of fermions hopping on the square-octagon lattice which consists of a square lattice with plaquette corners themselves decorated by squares.  Upon the inclusion of second neighbor spin-orbit coupling or non-Abelian gauge fields, time-reversal symmetric topological $Z_2$ band insulators are realized.   Additional insulating and gapless phases are also realized via the non-Abelian gauge fields. Some of the phase transitions involve topological changes to the Fermi surface. The stability of the topological phases to various symmetry breaking terms is investiaged via the entanglement spectrum.  Our results enlarge the number of known exactly solvable models of $Z_2$ band insulators, and are potentially relevant to the realization and identification of topological phases in both the solid state and cold atomic gases.
\end{abstract}
\date{\today}

\pacs{71.10.Fd,71.10.Pm,73.20.-r}


\maketitle

\section{Introduction \label{introduction}}

Topological phases of matter have recently been the focus of intense
theoretical and experimental effort.  Notable among them are the integer
and fractional quantum Hall liquids,\cite{Wen,Nayak:rmp08} which cannot be understood
in terms of the traditional description of phases based on symmetry breaking and local order parameters. While the quantum Hall states experimentally arise under strong magnetic fields, a new paradigm of topological phases\cite{Volovik} has emerged:  the so-called topological band insulators (TBI) which occur in the presence of time-reversal symmetry and spin-orbit coupling.\cite{Moore:nat10,Hasan:rmp10,Qi:pt10}  

The TBIs are similar to the quantum Hall states in that they possess a gapped bulk spectrum and gapless edge states.  However, a remarkable feature of TBI is that they can occur in both two\cite{Kane1,Kane2,Bernevig:prl06} and three spatial dimensions\cite{Moore:prb07,Fu:prl07,Roy:prb09} (while quantum Hall states are restricted to two dimensions).  Following the initial predictions of a two dimensional TBI in HgTe quantum wells,\cite{Bernevig:sci06} experiment revealed this intriguing quantum phase of matter.\cite{Konig:sci07,Roth:sci09}  Not long afterwards, predictions were made for three dimensional compounds \cite{Teo:prb08,Zhang:np09,Chadov:nat10,Lin:cm10} and a number of these have now been verified experimentally.\cite{Hsieh:nat08,Hsieh:sci09,Xia:np09,Chen:sci09,Hsieh:nat09,Hsieh:prl09}  Thus, TBI are now an experimentally established quantum state of matter.

The classification\cite{Schnyder:prb08} of TBI is based on a $Z_2$ invariant\cite{Kane1,Kane2,Moore:prb07,Fu:prb07,Fu:prl07,Roy:prb09} (or invariants, depending on the spatial dimension), rather than the Chern number used in the integer quantum Hall systems.  The $Z_2$ number turns out to be related to the parity of the number of gapless edge modes (Dirac nodes) appearing on the surface of an insulator:  Any odd number is topologically non-trivial and classifies the insulator as a TBI, while any even number can be shown to be adiabatically connected to the case with no gapless edge modes.   For a band insulator, the $Z_2$ number can be directly computed from the band structure.\cite{Kane1,Kane2,Fu:prb07,Fu:prl07,Teo:prb08}  With a straightforward procedure in hand to classify band insulators as topological or ``trivial", the search is on to determine which models, and therefore which physical systems, are expected to reveal TBI physics.   

Clearly, real systems involve electron-electron interactions, but in many cases these can be treated accurately within a mean-field approximation\cite{Bernevig:sci06,Teo:prb08,Raghu:prl08,Zhang:prb09,Zhang:np09,Chadov:nat10,Lin:cm10} in which case the relevant physics comes down to single-particle band physics.  Therefore, it is important to understand which features of a system lead to topological properties in the band structure, whether that band structure is derived from a non-interacting model or results from the self-consistent treatment of an interacting problem.  We emphasize that even in the absence of ``microscopic" spin-orbit coupling, TBI can result in interacting systems at the mean-field level by spontaneously generated spin-orbit coupling.\cite{Raghu:prl08,Zhang:prb09,Wen:prb10}  We also note that TBI have the ``convenient" property that both interactions of moderate strength\cite{Pesin:np10,Rachel10} and disorder\cite{Jain:prl09,Groth:prl09} can sometimes enlarge the region of parameter space (for fixed spin-orbit coupling) where the topological phases exist, thus aiding their realization in experiment.

\begin{figure}
\begin{center}
\includegraphics[width=8cm]{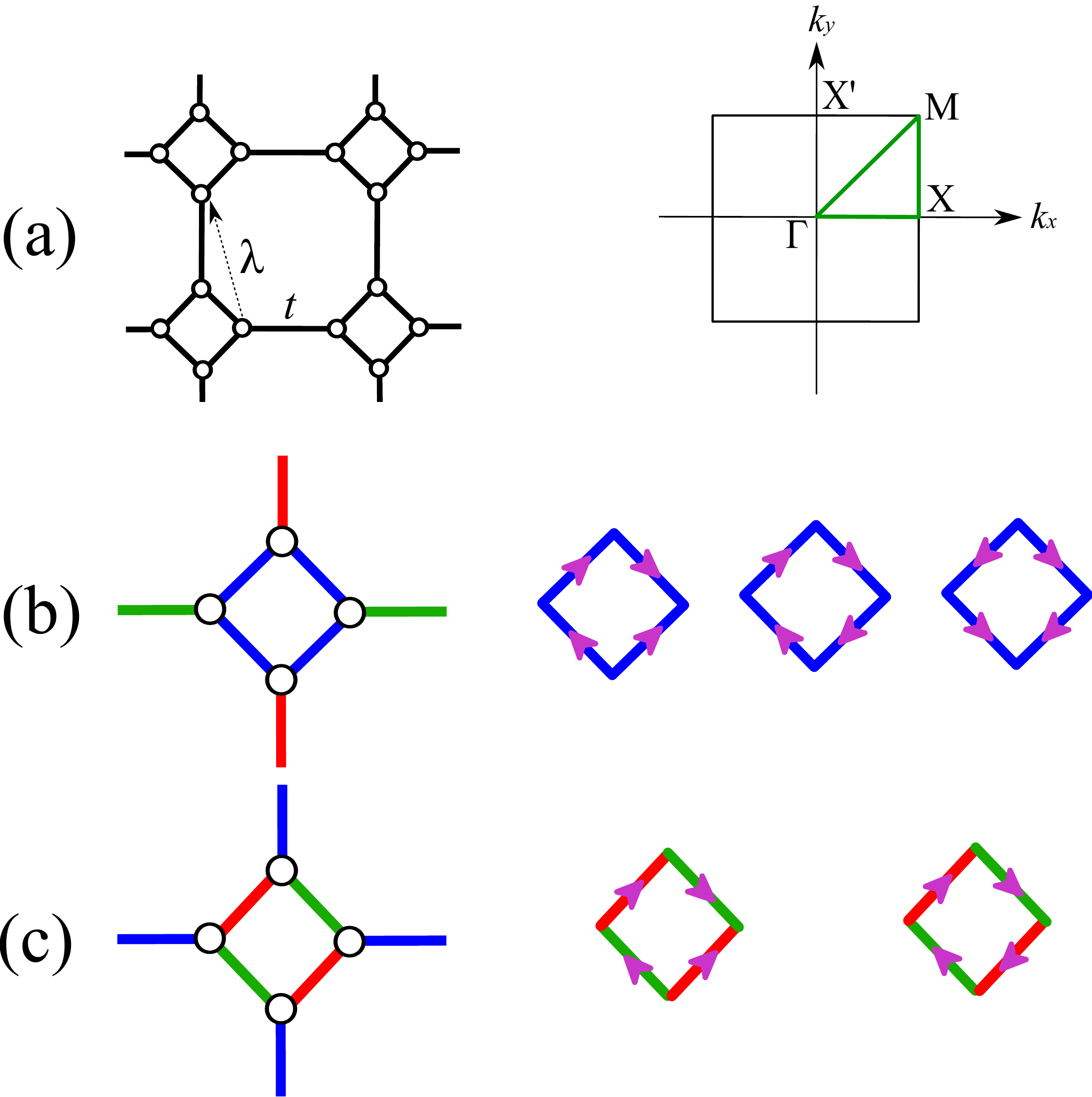}
\caption{(color online) (a) (left) Schematic representation of square-octagon lattice with inter-square hopping $t$ and second neighbor hopping $\lambda$ shown(dashed arrow line). Each vertex square and ``half" its emanating links is a unit cell of the lattice. (right) The Brillouin zone of the square-octagon lattice which has cubic symmetry (with a 4-point basis).  High symmetry points $\Gamma, \mathrm{M}, \mathrm{X}$ and $\mathrm{X'}$ are labelled.  (b,c) Illustration of gauge fields on nearest-neighbor hopping links. Different colors identify the way in which the gauge fields are applied:  Blue (dark gray), green (light gray) and, red (gray) correspond to $U_z$,$U_x$ and $U_y$ gauge fields (as defined in the text), respectively. For each case a variety of inversion symmetric  and asymmetric configurations are shown on the  right. Arrows on each link indicate the sign of the gauge field. } \label{lattice}
\end{center}
\end{figure}

In this paper, we focus on a particular two dimensional system:  a tight-binding model on the square-octagon lattice illustrated in Fig.\ref{lattice} that we show exhibits a number of $Z_2$ topological phases. The square-octagon lattice consists of a square lattice with squares placed at the corner of each square plaquette in a square lattice.  Part of the motivation to study this particular lattice comes from the knowledge that it supports topological phases in spin models, with the celebrated toric code\cite{toric} as an effective low energy description.  For certain parameter regimes in the spin models, Abelian and non-Abelian phases are obtained,\cite{mosaic} as well as a pseudo-Fermi surface.\cite{baskaran}  The connection between topological spin systems and topological band insulators was recently explored on the decorated honeycomb lattice in Ref.[\onlinecite{Ruegg:prb10}].  We find that the square-octagon lattice tight-binding model of fermions we study indeed realizes a TBI phase over a wide range of parameters, and there are a number of interesting quantum phase transitions appearing in the phase diagram.

With this work, we expand the number of known lattices (honeycomb,\cite{Kane1,Kane2} kagome,\cite{Guo:prb09} checkerboard,\cite{Sun:prl09} decorated honeycomb,\cite{Ruegg:prb10} Lieb,\cite{Weeks:10} pyrochlore,\cite{Guo:prl09} perovskite,\cite{Weeks:10} and diamond\cite{Fu:prl07,Zhang:prb09}) that support topological phases within a simple s-wave tight-binding approximation.  When more orbitals are included, such as those with $p$ and $d$ character, the number of lattices supporting topological phases is even larger.\cite{Bernevig:prl06,Pesin:np10,Yang:10}  
Our main goal in this work is to help identify the conditions required to obtain topological phases, and then predict how that physics can be observed in experiment.  Our results are relevant to both solid state and cold atom systems.

This paper is organized as follows.  In Sec.\ref{model} we introduce a single-orbital model of fermions on the square-octagon lattice.  We show topological phases are realized upon the inclusion of a second-neighbor spin-orbit coupling term similar to that introduced by Kane and Mele\cite{Kane1,Kane2}.  In Sec. \ref{non-Abelian} we show that in the absence of second-neighbor hopping, topological phases can be realized with non-Abelian gauge fields placed on the links of the lattice.  As a function of the strength and type of non-Abelian gauge fields on the links,  we find an exceptionally rich phase diagram and study it in some detail.  Finally, in Sec. \ref{sec:entanglement} we study the stability of the topological phases to various symmetry breaking terms using the entanglement spectrum.  In Sec.\ref{conclusions} we summarize the main results of this work.

\section{ Kane-Mele type Model Hamiltonian\label{model}}
 \label{model}
The lattice we consider in this paper is shown in
Fig.\ref{lattice} (a). We first study a Kane-Mele type model Hamiltonian,\cite{Kane1,Kane2}  
\bea \label{squareH} H=H_0+H_{SO},\eea 
where
 \bea
\label{H0} H_0=-\sum_{i,j\in
\diamondsuit,\sigma}c^{\dag}_{i\sigma}c_{j\sigma}-t\sum_{\diamondsuit\rightarrow\diamondsuit,\sigma}c^{\dag}_{i\sigma}c_{j\sigma},\eea
and
\bea
\label{eq:HSO}
H_{SO}=i\lambda\sum_{\ll
i,j\gg}c^{\dagger}_{i\alpha}\left(\overrightarrow{e}_{ij}.\overrightarrow{\sigma}\right)_{\alpha \beta}c_{j\beta}.
\eea
The first and second terms in $H_0$ describe the hopping of fermions with spin $\sigma$ on
square plaquettes and between neighboring plaquettes, respectively.  We have set the nearest-neighbor on-plaquette hopping amplitude to unity and expressed all other energies in terms of this.  Here $c^{\dag}_{i\sigma}$ creates a fermion of spin $\sigma$ on site $i$ and $c_{i\sigma}$ annihilates a fermion of spin $\sigma$ on site $i$. The Hamiltonian $H_{SO}$ describes the spin-orbit coupling between second nearest sites whose relative position is encoded in the unit vector $\overrightarrow{e}_{ij}$, using the usual conventions: $\vec e_{ij}=\frac{\vec d^1_{ij} \times \vec d^2_{ij}}{|\vec d^1_{ij} \times \vec d^2_{ij}|}$, the vector $\vec d^1_{ij}$ points from site $j$ to a nearest neighbor site to both it and site $i$, and $\vec d^2_{ij}$ points from that nearest neighbor to the site $i$ which is a second neighbor to site $j$.\cite{Kane1,Kane2,Zhang:prb09,Essin:prb07} 

\begin{figure}
\begin{center}
\includegraphics[width=7cm]{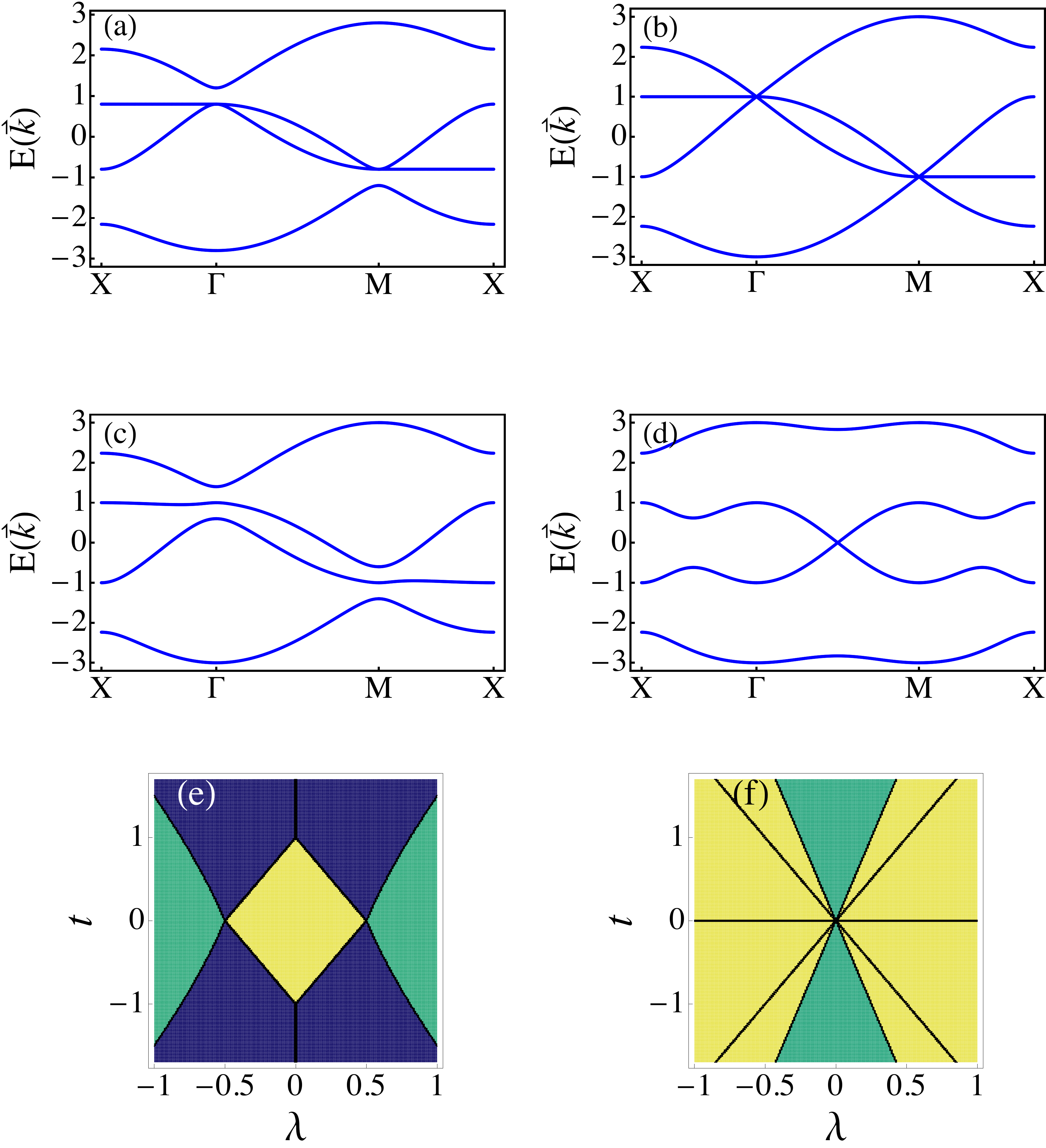}
\caption{(color online) Two top panels: the band structure of the tight-binding 
square-octagon lattice model, Eq.\eqref{squareH}, along various high symmetry directions 
for the path shown in Fig.\ref{lattice}(a).  Parameters are: (a) $t=0.8, \lambda=0$, (b) $t=1, \lambda=0$, (c) $t=1, \lambda=0.1$ and (d) $t=1, \lambda=0.5$.
Bottom panel: phase diagram of the model at filling factor (e) $1/4$ and (f) $1/2$. Different phases are distinguished by colors as follows. blue (black): topological band insulator (TBI) state, yellow (light): band insulator (BI)  and green (gray): semimetal (SM). Note that at 1/2 filing, there is no QSH state in this model and at 1/4 filling larger $t$ tends to stabilize the topological phase.  In (e) the solid vertical line $\lambda=0$ is a semi-metal. The two intersecting dark lines in (f) running through the BI phases are lines where the system is a semi-metal with 4 Dirac points in the band structure as in Fig.\ref{bands}(d).} \label{bands}
\end{center}
\end{figure}

Exploiting the translational symmetry of the model, the Hamiltonian can be diagonalized as
$H=\sum_{\mathbf{k}\sigma}\Psi^{\dagger}_{\mathbf{k}\sigma}\widetilde{H}_{\mathbf{k}\sigma}\Psi_{\mathbf{k}\sigma}$,
where
$\Psi^{\dagger}_{\mathbf{k}\sigma}=(c^{\dagger}_{1\mathbf{k}\sigma},c^{\dagger}_{2\mathbf{k}\sigma},c^{\dagger}_{3\mathbf{k}\sigma},c^{\dagger}_{4\mathbf{k}\sigma})$
represents  the four sites around the square unit cell (those that sit at the sites of the underlying square lattice), and $\widetilde{H}_{\mathbf{k}\sigma}$ is the Hamiltonian in Fourier space. The
band structure for different values of $t$ and $\lambda$ of the model along the directions of high symmetry in the Brillouin zone is shown in Fig.\ref{bands}(a-d). Each band is doubly degenerate due to the spin degree of freedom, and some bands touch each other at different crystal momenta in the Brillouin zone. For $t \neq 1$ the behavior is shown in Fig.\ref{bands}(a), and for $\lambda=0$ we have a band touching at a single point, where a quadratic band touches a locally flat band, similar to what occurs in the kagome lattice and decorated honeycomb lattice.\cite{Guo:prb09,Ruegg:prb10}
An interesting feature of this lattice is that there are three bands (6 including spin degeneracy) that cross at the $\Gamma$ and $\mathrm{M}$ points for $t=1$ and $\lambda=0$ (see Fig.\ref{bands}(b)), which are time reversal invariant momenta of the square lattice. At these points a locally flat band meets a Dirac-like structure coming from the other two bands. The situation is very similar to what occurs in the decorated honeycomb lattice model at hopping parameters $t'/t=1.5$ where a Dirac point is intersected by a flat band.\cite{Ruegg:prb10}  

In the present work we are interested in insulating phases, and one can follow
different routes to gap the band structure of this model. One route
is to localize the electrons to isolated square plaquettes. This is
done by decreasing the hopping $t$ in $H_0$. As $t \to 0$ this model
becomes gapped at 1/4 filling and 3/4 filling as shown in Fig.\ref{bands}(a). It is trivial that at
the extreme limit $t=0$, the lattice becomes a set of disconnected
plaquettes and is thus a trivial, non-topological insulator.  All insulating phases created in this
way  are continuously (adiabatically) connected to this insulator. We therefore do not study this limit of
Eq.\eqref{H0} because this evidently gives rise to trivial insulators which are not the focus of this paper. The same trivial insulator phase also appears in the
Kagome lattice by introducing a pattern of alternating bonds.\cite{Guo} 

Another approach to opening a gap is based on including a spin-orbit coupling term, Eq.\eqref{eq:HSO}, that changes the semi-metallic state of the model at 1/4 and 3/4 filling into an insulator as shown in Fig.\ref{bands}(c).  From a low-energy analysis around the Dirac+flat band crossing points, one can deduce that a gap of value $\Delta=|4\lambda+2t-2|$ (note for $t=1$ this vanishes for $\lambda=0$) opens up in the presence of spin-orbit coupling. However, at 1/2 filling, though the second and third bands separate from one another, for a wide range of spin-orbit coupling the Fermi energy crosses the bands resulting in a semimetallic state. In particular, at coupling $|\lambda/t|=0.5$, a Dirac crossing occurs at crystal
momenta $(\pm\frac{\pi}{2},\pm\frac{\pi}{2})$ [see Fig.\ref{bands}(d)], and upon further increasing of the coupling $|\lambda|$ a gap develops. We will further discuss 
the implication of these Dirac nodes below.

\begin{figure}
\begin{center}
\includegraphics[width=8cm]{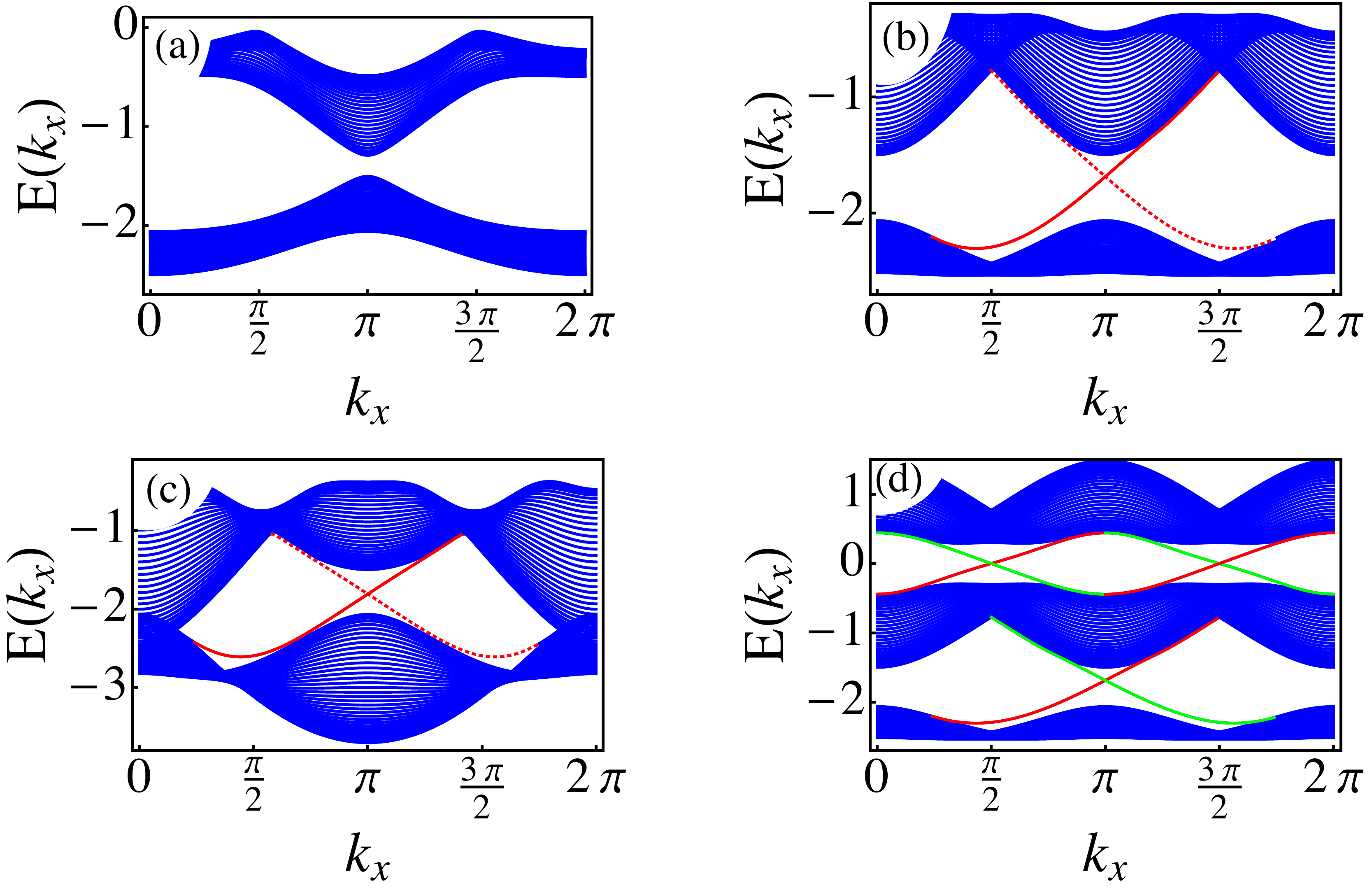}
\caption{(color online) Energy dispersion of Eq.\eqref{squareH} in a strip geometry at different spin-orbit
coupling $\lambda$ for fixed $t=0.5$ and 1/4 filling. Shown are the cases: (a) $\lambda=0.2$ (b) $\lambda=0.5$ (c) $\lambda=0.8$.  The states crossing the gap are edge modes, and on each edge of the strip right (solid red line) and left moving (dashed red line) states are degenerate at a single time-reverasal invariant momentum, which is a signature of a $Z_2$ topological band insulator.  Directly computing the $Z_2$ invariant indeed proves that (b) is a topological insulator  state.  In (c) there at also edge modes present, but the bulk is not insulating. The situation is reminiscent of that found in pure Sb.\cite{Fu:prb07,Teo:prb08} In (d) we are considering a Haldane type model breaking the time reversal symmetry by adapting a spinless model using the same parameters as in (b). The red (dark) and green (grey) lines stand for edge modes at right and left sides of the strip, respectively.  The absolute value of the Chern number is 1 at 1/4 filling and 2 at 1/2 filling.}\label{fig:strip_Kane}
\end{center}
\end{figure}

The gapped bulk phases at different filling fractions may be topologically trivial or nontrivial, and this is most easily seen via the edge modes:\cite{Kane1,Kane2} the former is characterized by even number of Kramers pairs, while the latter is distinguished by an odd number of Kramers pairs.  Because our model has inversion symmetry, the $Z_2$ invariants can be easily read off from the parity eigenvalues of the wavefunction at different time reversal momenta.\cite{Fu:prb07} Fig.\ref{bands}(e) and (f) depicts how competition between parameters in the Hamiltonian drives the model into different phases at 1/4 and 1/2 filling. The case of 3/4 filling results in the same phase diagram as 1/4 filling because of the particle-hole symmetry present in the model (with $\lambda=0$).

As discussed above, the topological insulator supports helical edge states at its boundary. To calculate the edge spectrum, we have considered our model on a strip geometry which allows us to see the edge-projected momenta which clearly reveal edge states crossing the gap.  Fig. \ref{fig:strip_Kane} shows
the spectrum at 1/4 filling for different values of spin-orbit coupling corresponding to the different phases in Fig.\ref{bands}(e). Fig. \ref{fig:strip_Kane} shows the behavior of edge modes through the transition between different phases. In the trivial insulator phase there are no gapless edge modes, effectively merged with the gapped bulk modes [see Fig.\ref{fig:strip_Kane}(a)].  However, as expected in the
topological insulator phase [see Fig.\ref{fig:strip_Kane}(b)], the edge modes traverse the bulk band gap. Importantly, there is an odd number of edge modes, which is the source of topologically robustness of these edge modes to weak disorder. Focusing at one of the edges of the strip, we can see that two traversing edge modes (solid and dashed red lines in Fig.\ref{fig:strip_Kane}(b)) are a Kramers pair crossing each other at time reversal invariant momentum $\pi$. The final figure of this sequence, Fig. \ref{fig:strip_Kane}(c), is in a metallic phase, which is clear since the system is gapless.  Some values of the edge-projected momenta have bulk conduction and valence bands merged into one another. However, even in this case we can see that there are traversing modes around the time-reversal invariant point $\pi$.  This mode can be shown to be localized on the edge of the system and behaves in many ways like the edge modes of the topological insulator phase.  However, disorder on the edge can mix this state with bulk states, so it is not ``topologically protected".\cite{Fu:prb07,Teo:prb08}  The physics of such boundary states in the presence of disorder has been investigated in Ref.[\onlinecite{Refael}].  In closing this part, we note that the stability of the topological phases to Rashba coupling and charge density wave order imposed via a staggered on-site potential is very similar to that found for the honeycomb lattice\cite{Kane2} and decorated honeycomb lattice.\cite{Ruegg:prb10}  We will investigate these effects in more detail in Sec.\ref{sec:entanglement} via the entanglement spectrum.

As we discussed earlier, the trivial insulting phase can not be adiabatically connected to a topological insulating phase; along the way the gap $\Delta=0$ must be closed. In this model, the low energy physics close to the gap closing point can be described by Dirac fermions. The corresponding Dirac node appear at the $\mathrm{M}$ point. At the phase transition the sign of the mass gap changes.\cite{Bernevig:prl06,Bernevig:sci06,Murakami:prb07} Although this latter point often is a common feature of topological insulators and Hall systems which, respectively, preserve and break time reversal symmetry, one may wonder about the presence of only a single Dirac node. According to the Dirac low energy theory, each Dirac node should contribute half of the quantized Hall current, namely $e^2/2h ~ \mathrm{sgn(m)}$, to the total Hall current at the edge of the system,\cite{Oshikawa:prb94} where $\mathrm{sgn(m)}$ is the sign of the mass gap. Thus, a naive expectation results in a half-odd integer when there are an odd number of Dirac fermions in the system, which contradicts the integer quantization of the Hall conductance as given by TKNN integer.\cite{tknn:prl82} In fact, the total contribution to the Hall conductance is not given by only summing up the individual contribution of Dirac fermions. Indeed, the TKNN integer (or Chern number) carries information regarding the Bloch states through the entire Brilloun zone, not just around the Dirac nodes.  Thus, the vorticity content of the Bloch states over the entire Brillouin zone is important.\cite{Hatsugai:prb96} However, the number of Dirac nodes gives the correct {\em change} of the Hall conductance through a gap closing process.\cite{Oshikawa:prb94}  

For a time reversal invariant system the Chern number is zero. To get a nonzero value, one has to break time reversal symmetry. Here, we do this by considering only one spin species, say up, which is a kind of Haldane model\cite{Haldane:prl88} on the square-octagon lattice. The bulk and edge modes at $1/4$ and $1/2$ filling for a strip geometry is shown in Fig. \ref{fig:strip_Kane}(d). The solid red (dark grey) and green (grey) lines traversing gap stand for edge modes localized at right and left of the strip, respectively. At $1/4$ filling the Chern number in $n=1$ which is consistent with the number of modes per edge.\cite{Hatsugai:prl93}  At this filling via the gap closing point the Chern number changes from $n=0$ to $n=1$, being equal to the number of Dirac nodes. 

As we have seen in Fig.\ref{bands}(f), at $1/2$ filling there is no topological insulating phase. We can see that two $Z_2$ trivial insulator phases are separated by a gap closing point along two  intersecting dark lines. At each gap closing point, four Dirac nodes appear in the Brilloun zone [see Fig.\ref{bands}(d)]. For these trivial insulator phases, the Haldane type model considered shows the Chern number is not zero. Instead, it is $n=2$ before and $n=-2$ after the gap closing point, which correspond to having two edge modes in the bulk gap as shown in Fig. \ref{fig:strip_Kane}(d) (two red and two green lines). Thus, the Chern number, or alternatively the anomalous Hall current, will change by 4 through the gap closing point. Once more we see that the {\em change} of the Chern number coincides with the number of Dirac nodes.

\section{Hamiltonians with non-Abelian gauge fields: Topological Insulators and Topological Phase Transitions}
\label{non-Abelian}

In the preceding section we used second neighbor spin-orbit coupling to induce a topological insulator phase in a tight-binding model, which is a well-known paradigm for driving a transition to a topological insulator phase.\cite{Kane1,Kane2,Guo:prb09,Sun:prl09,Ruegg:prb10,Weeks:10,Guo:prl09,Fu:prl07,Zhang:prb09}
Indeed, the essential ingredient comes from the spin-dependent gauge fields\cite{Shitade:prl09} (i.e., the spin-dependent second neighbor hopping in the most familiar cases). In this section we introduce a set of gauge fields living on the links between nearest-neighbor sites of the lattice. These gauge fields can be artificially induced\cite{Osterloh,Lin:prl09,Spielman:pra09} in the many-body Hamiltonian of ultracold atoms in optical lattices, where many rich behaviors can and haven been explored.\cite{Bloch} For instance, ultracold atoms trapped in a honeycomb lattice, when subjected to non-Abelian gauge fields, exhibit various phases possessing different quantum orders\cite{miguel1} due to the coupling of the gauge fields to the emergent relativistic quasiparticles. Here we address how such gauge 
fields may give rise to different phases of the model Eq.\eqref{squareH} with only nearest-neighbor hopping. In particular, we are interested in whether such fields can stabilize a $Z_2$ topological insulator phase.\cite{Goldman:cm10,Bermudez:cm10}   Below we show that applying gauge fields to nearest-neighbor links leads to $Z_2$  topological band insulator phases as well as topological changes in the Fermi surface in the metallic phases of the model.  

We assume that the hopping terms in the first term of Eq.(\ref{H0}) are modified by some unitary matrices, $\hat{U}$, as,\cite{miguel1}
 \bea \label{gaugeH}
H=\sum_{<i\sigma,j\sigma'>}[U_{ij}]_{\sigma\sigma'}c^{\dag}_{i\sigma}c_{j\sigma'}+\mathrm{H.c.}
\eea
For spin 1/2 particles, a natural choice for unitary matrices will be the two-dimensional representation of the corresponding Lie group,
 \bea \label{unitary}
U_{z}=e^{i\gamma\sigma^z}, U_{x}=e^{i\alpha\sigma^x}
,U_{y}=e^{i\beta\sigma^y},
\eea 
where $\gamma,\alpha$ and $\beta$
are parameters related to gauge fluxes, and $\sigma^\nu$ for $(\nu=z,x,y)$ stands for the usual Pauli matrices. Note that these gauge fields preserve the time reversal symmetry, but may be applied 
in ways that could either preserve or break inversion symmetry. We consider two patterns for the modulation of hopping terms by applying gauge fields on the links as shown in Fig.\ref{lattice}(b,c).  These two patterns  give rise to rich phenomena that could potentially be realized in optical lattices.

\subsection{Gauge fields and topological insulating phases}

\begin{figure}
\begin{center}
\includegraphics[width=8cm]{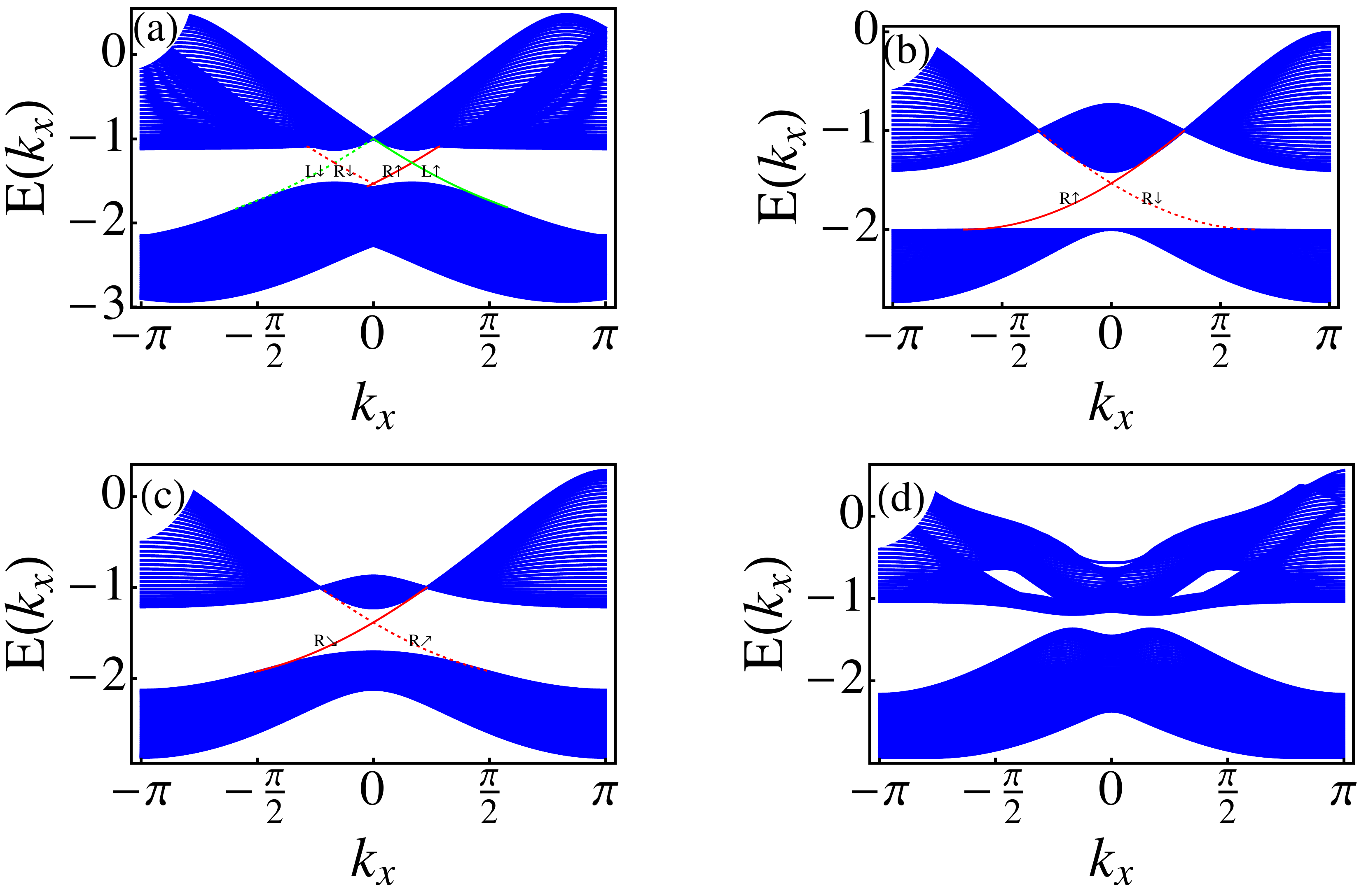}
\caption{(color online) Energy spectrum for the tight-binding model with nearest-neighbor gauge fields solved on the strip geometry.  The gauge field modulations corresponding to the patterns shown in Fig.\ref{lattice}(b,c) were used.  (a) shows the edge modes with inversion symmetry broken and (b) inversion symmetry preserved  using the gauge fields in Fig.\ref{lattice}(b) with parameters $\alpha=\beta=0$ and $\gamma=\pi/6$ . (c) shows the corresponding edge modes with inversion preserved and (d) inversion broken using the signs in  Fig.\ref{lattice}(c) with parameters $\alpha=\beta=\pi/6$ and $\gamma=0$. The red (solid and dashed) and green (solid and dashed) lines stand for right and left of the strip, respectively, and arrows indicate different spin orientations. By counting the parity of the edge modes, we see that (a), (b) and (c) are topologically non-trivial, while (d) is topologically trivial.} \label{edgegauge}
\end{center}
\end{figure}

We start by focusing on the pattern shown in Fig.\ref{lattice}(b).  First we set $\alpha=\beta=0$ implying 
that only hopping around square plaquettes on each point of the underlying square lattice have spin dependency. As depicted on the right hand of Fig.\ref{lattice}(b), the
gauge fields can be selected so that the inversion symmetry is
preserved or not. Three different patterns are depicted. In the first pattern the inversion symmetry is broken.
Here, the fluxes $-2\gamma (2\gamma)$ are attached to square (octagon) plaquettes, and the degeneracy between spin up and down is lifted.
Only at time reversal invariant momenta do the different spin states remain degenerate.  However,
any nonzero value of $\gamma$ creates a gap in the spectrum at 1/4
filling. This insulating phase is a topological band insulator as one pair
of edge modes appear at the edge of the model in a strip
geometry, as shown in Fig.\ref{edgegauge}(a). In this case we can determine the spin-dependency of each edge mode, since the gauge field $U_{z}$ conserves the spin up and down components. At each edge there in one Kramers pair crossing each other at time reversal momentum $k_x=0$. In this figure the right (R) and left (L) edge are distinguished by red and green colors, respectively, and $\uparrow (\downarrow)$ stands for up (down) spin orientation. 

The second pattern of gauge fields  in Fig.\ref{lattice}(b) preserves the inversion symmetry of the lattice, and the square (octagon) plaquettes carry $-4\gamma(4\gamma)$ fluxes. Different spin
projections will have the same energy.  Once again any nonzero value
of $\gamma$ (except at $\gamma=\pi/4$ where the flux pattern is
equivalent to the zero-flux pattern because of the particle-hole symmetry) opens a gap at 1/4
filling. The main difference between the inversion symmetric and non-inversion symmetric flux patterns appears in nature of the edge modes. As shown in Fig.\ref{edgegauge}(b), there is one pair of edge modes traversing gap showing that for gauge fields that preserve the inversion symmetry the insulating phase is a topological phase. Direct evaluation of the $Z_2$ invariant also demonstrates this.  When inversion symmetry is broken, a topological phase still results, as seen in Fig.\ref{edgegauge}(a).  However, the edge dispersion gets ``split" relative to the inversion symmetric case.

The third pattern of gauge fields  in Fig.\ref{lattice}(b) also preserves the inversion symmetry, but the
plaquettes no longer carry fluxes. The model remains in the semimetallic phase at
all fillings and for all values of $\gamma$, and behaves effectively as if no gauge
fields are applied. This corresponds to the canceling part of the 
second-neighbor spin-orbit coupling on the square plaquetttes. In fact, according to Eq.(\ref{squareH}), the only nonzero contribution of spin-orbit coupling comes from the 
hopping between squares, while the second-neighbor hopping on the squares 
cancel each other. 

We also examined the stability of the topological insulator phase by turning on the gauge fields applied to the links connecting squares [red and green links in Fig.\ref{lattice}(b)]. We found that as the value of either $\alpha$ or $\beta$ is increased, there is a critical value for which the gap
in the topological phase closes and therefore a quantum (topological) phase transition generally occurs.

\subsection{Topological phase transitions}
In this subsection we turn to a different method for modifying the hopping terms: applying the gauge fields as shown in Fig.\ref{lattice}(c). We assume
that only the hopping terms around square plaquettes at the sites of the underlying square lattice undergo such a modulation by the nontrivial gauge fields and we set $\gamma=0$. Here we can also  consider different configurations respecting or violating inversion symmetry. 
Let us first consider the case in which the inversion symmetry is preserved. Applying gauge fields with finite $\alpha$ and $\beta$  gaps the spectrum at 1/4 and 3/4 fillings. Again, we find that
 at 1/4 filling  the gapped phase is topological. This fact is readily deduced by looking at 
the spectrum of the edge modes when the model is solved on a strip geometry, as seen in Fig.\ref{edgegauge}(c).  The nonzero values of $\alpha$ and $\beta$ breaks the symmetry between up and down spins, and thus right or left movers are mixing states of different spin orientations. A direct evaluation of the $Z_2$ invariant also shows this is a TBI.  We note that at $\alpha=\beta=\pi/2$ the gap in the spectrum at 1/4 (and 3/4) filling vanishes and further increasing $\alpha$ and $\beta$  opens it again and returns the model to a topological insulating phase.

\begin{figure}
\begin{center}
\includegraphics[width=8cm]{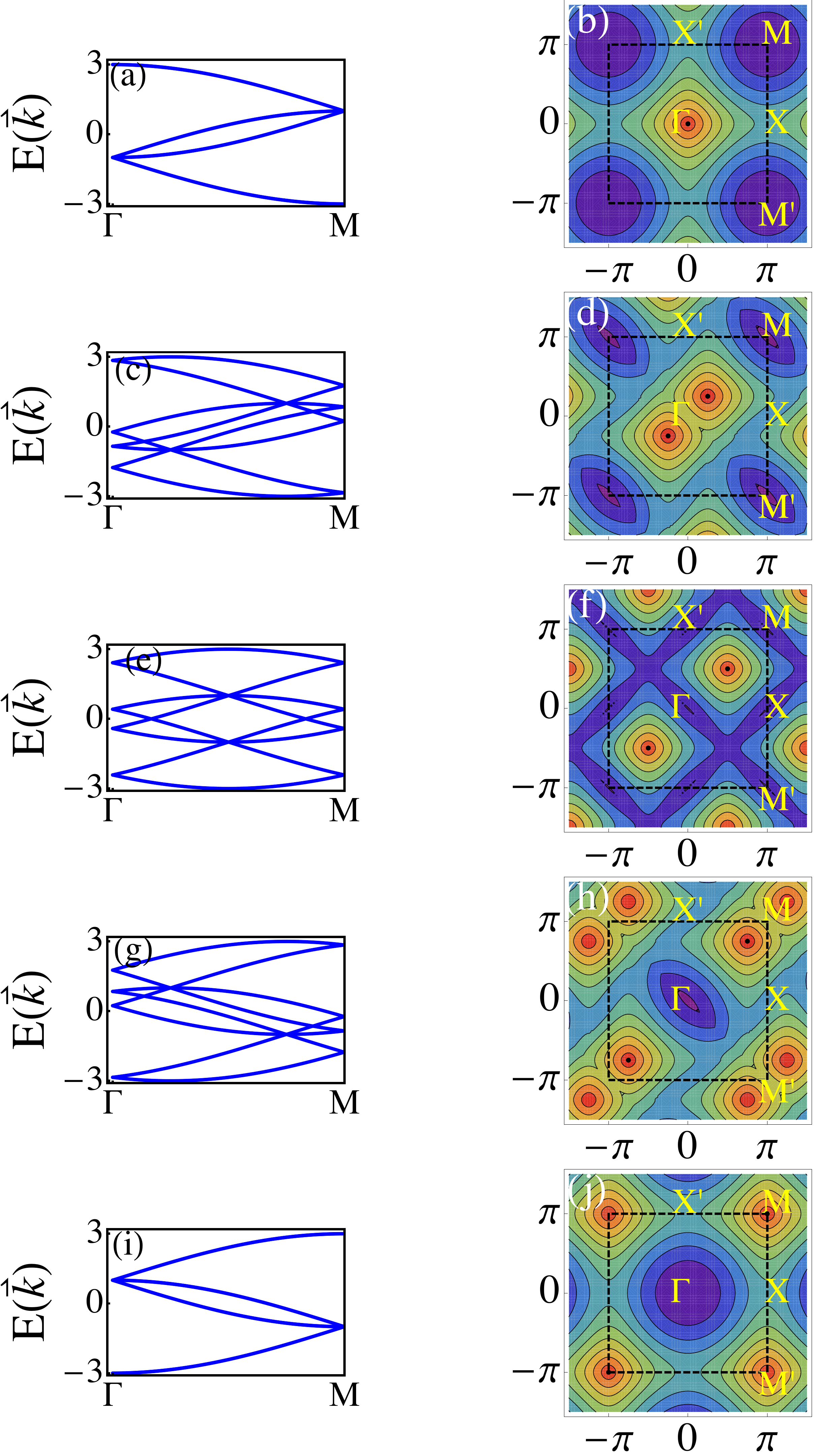}
\caption{(color online) Band dispersions (left) along $\Gamma$ to $\mathrm{M}$ and corresponding contour plot (right) over the whole Brillouin zone (dashed square) for $\alpha=\gamma=0$: in  (a,b) $\beta=0$, in (c,d) $\beta=\pi/4$, in (e,f) $\beta=\pi/2$, in (g,h) $\beta=3\pi/4$, in (i,j) $\beta=\pi$. Note that Dirac points move across the Brillouin zone as $\beta$ evolves.  The Dirac points are always intersected by a locally flat band. } \label{gauge1}
\end{center}
\end{figure}

A richer set of behaviors is observed when we consider the case where the gauge fields in Fig.\ref{lattice}(c) break inversion symmetry.  As we discussed before, there are two Dirac
nodes (including spin) at 1/4 filling (when $t=1$). Upon the inclusion of
gauge fields, this degeneracy is split. To see this, let us
consider the  case of $\alpha=0$. Fig. \ref{gauge1} shows a set
of band dispersions (left) alongside constant energy contours showing the evolution 
of the Dirac nodes within the Brillouin zone (right). This set of gauge fields corresponds to the left side of the phase diagram shown in Fig.\ref{phasediagram} (cyan dashed line $\mathrm{L_1}$), and the stars are different values of $\beta$ in Fig. \ref{gauge1}.  Note that we have considered the band dispersions
along the path connecting $\Gamma$ to $\mathrm{M}$. In the absence of gauge fields there is single Dirac node (doubly degenerate) at the center of the
Brillouin zone, as seen in Fig.\ref{gauge1}(a) and (b). As the gauge field
$\beta$ increases from zero, two initially degenerate Dirac nodes are
split apart and start to move along the path connecting center of the
Brillouin zone to the its corner ($\Gamma$ to $\mathrm{M}$). These Dirac nodes are also intersected locally by a flat band. In the
language of Ref.[\onlinecite{miguel1}], there is no quantum phase
transition along this path as the total number of the Dirac nodes remains unchanged. They are only split and move about in the Brillouin zone. At the end of the path with $\beta=\pi$ [see Fig.\ref{gauge1}(i)] the two Dirac nodes are again merged into a single one at the corner of the Brillouin zone. We could consider any border of the phase digram in Fig.\ref{phasediagram}, they will present the same evolution. The scenario can be quite different if the gauge fields are applied in another manner.

\begin{figure}
\begin{center}
\includegraphics[width=8cm]{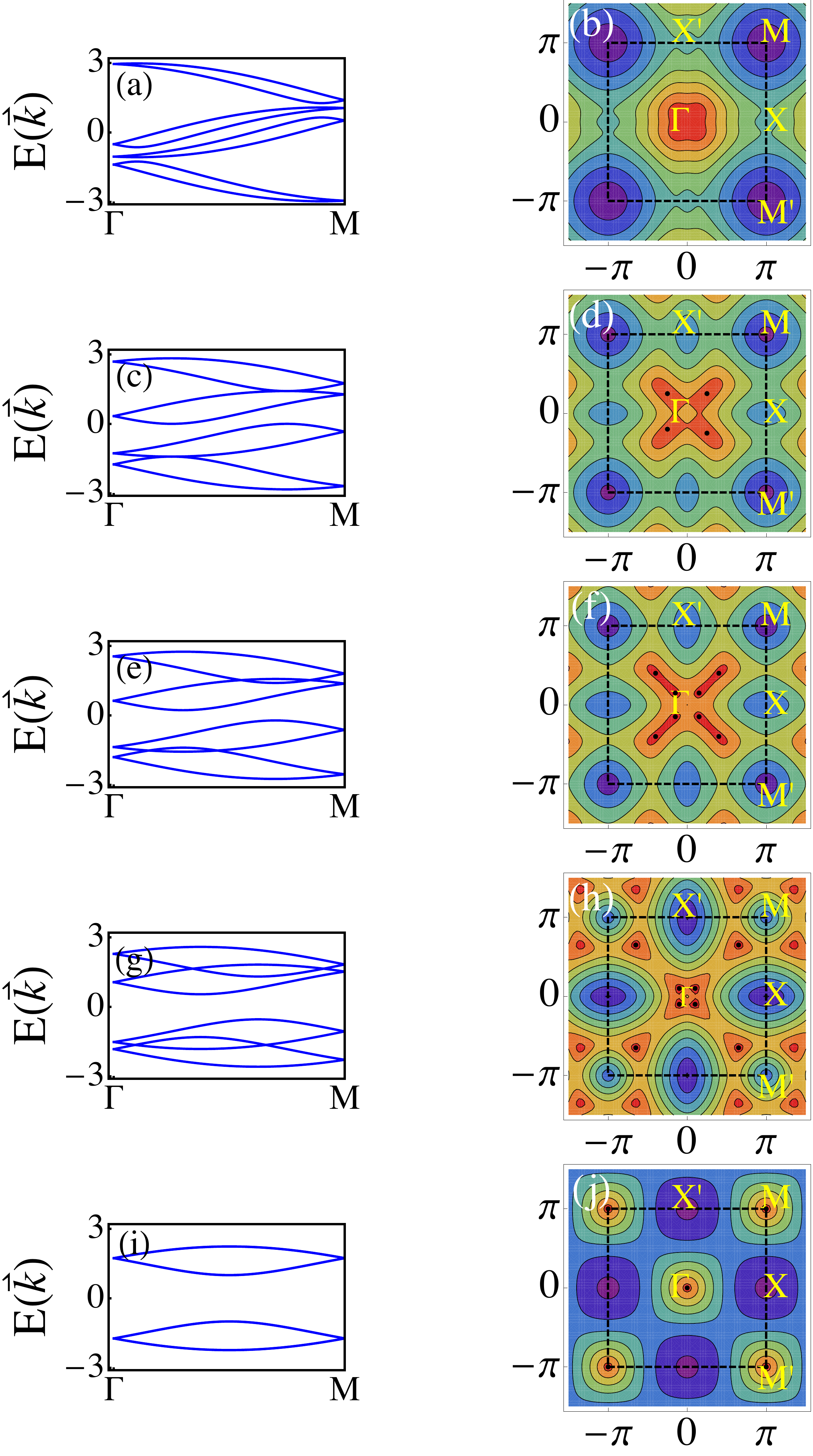}
\caption{(color online)  Band dispersions (left) along $\Gamma$ to $\mathrm{M}$ and corresponding contour plot (right) over the whole Brillouin zone(dashed square) for $\gamma=0$: in (a,b) $\alpha=\beta=0.1\pi$, in (c,d) $\alpha=\beta=\pi/4$, in (e,f) $\alpha=\beta=0.3\pi$, in (g,h) $\alpha=\beta=3\pi/8$, in (i,j) $\alpha=\beta=\pi/2$.  Note the quadratic band touching point in (c) splits into 2 Dirac points in (e), (g), and (i).} \label{gauge2}
\end{center}
\end{figure}

We now turn on both gauge fields $\alpha$ and $\beta$ with $\alpha=\beta$ [line $\mathrm{L_2}$ in Fig.\ref{phasediagram}], still corresponding to the inversion asymmetric configuration in Fig.\ref{lattice}(c) (because we already showed the inversion symmetric case results in a $Z_2$ TBI when the spectrum is gapped).  As seen in Fig.\ref{gauge2}(a) and (b), the value $\alpha=\beta=0.1\pi$ leads to a global gap at 1/4 filling. The insulating phase here is a trivial insulating phase since, as shown in Fig.\ref{edgegauge}(d), there are no states crossing the gap.  However, further increasing of $\alpha=\beta$ closes the gap at $\alpha=\beta=\pi/4$ [see Fig.\ref{gauge2}(c) and (d)] and a quadratic band touching point (QBTP)\cite{Sun:prl09} emerges along the line connecting $\Gamma$ to $\mathrm{M}$. In fact, in the parameter space spanned by $\alpha$ and $\beta$, as shown in Fig.\ref{phasediagram}, there are 4 such QBTP indicated by solid circles.  Since the gap is closed, we expect a quantum phase transition occurs at this point.\cite{Murakami:prb07} The nature of this transition is characterized by fine-tuning the gauge fields around the transition point. Fig. \ref{gauge2}(e) and (f) reveal how bands cross each other (as $\alpha=\beta$ is increase from the value of $\pi/4$)  giving rise to a pair of Dirac nodes in the low energy dispersion along $\Gamma$ to $\mathrm{M}$. Thus, 8
Dirac nodes appear (in the low-energy sector at 1/4 filling) in the Brillouin zone. Although part of the physics of the creation of Dirac nodes is similar to the non-Abelian hexagonal lattice,\cite{miguel1} the Dirac nodes here are created from a global vacuum.  Once the nodes are created, they move away from each other towards the high symmetry points of the lattice, which are the $\Gamma$ and $\mathrm{M}$ points. At $\alpha=\beta=\pi/2$ four Dirac nodes at $\Gamma$ and
$\mathrm{M}$ merge into a single one. Thus, at this point we have two Dirac nods in the Brillouin zone as indicated by $O$ in Fig.\ref{phasediagram}. As the gauge fields are increased further a converse process occurs. That is, the Dirac nodes at these points are split into four nodes and
eventually at $\alpha=\beta=3\pi/4$ are annihilated into vacuum. 

\begin{figure}
\begin{center}
\includegraphics[width=8cm]{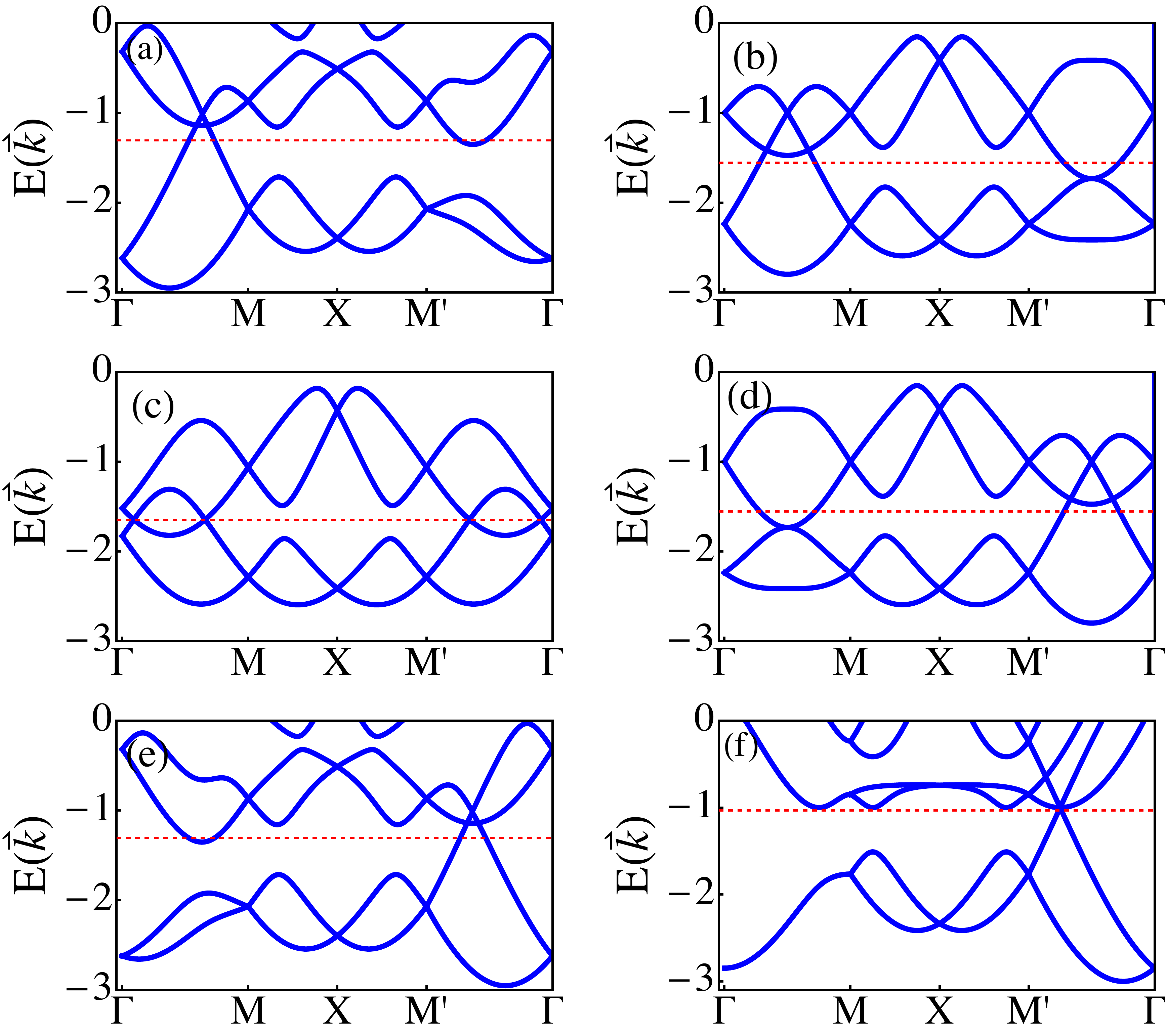}
\caption{(color online)  Band dispersions along the high symmetry points of the Brillouin zone at different gauge values for inversion symmetry breaking flux patterns with $\beta=-\alpha+3\pi/4$: in (a) $\alpha=\pi/8$, in (b) $\alpha=\pi/4$, in (c) $\alpha=3\pi/8$, in (d) $\alpha=\pi/2$, in (e) $\alpha=5\pi/8$, and in (f) $\alpha=3\pi/4$. Along the evolution of $\alpha$ fermi pockets appear and disappear.} \label{phase}
\end{center}
\end{figure}

\begin{figure}
\begin{center}
\includegraphics[width=8cm]{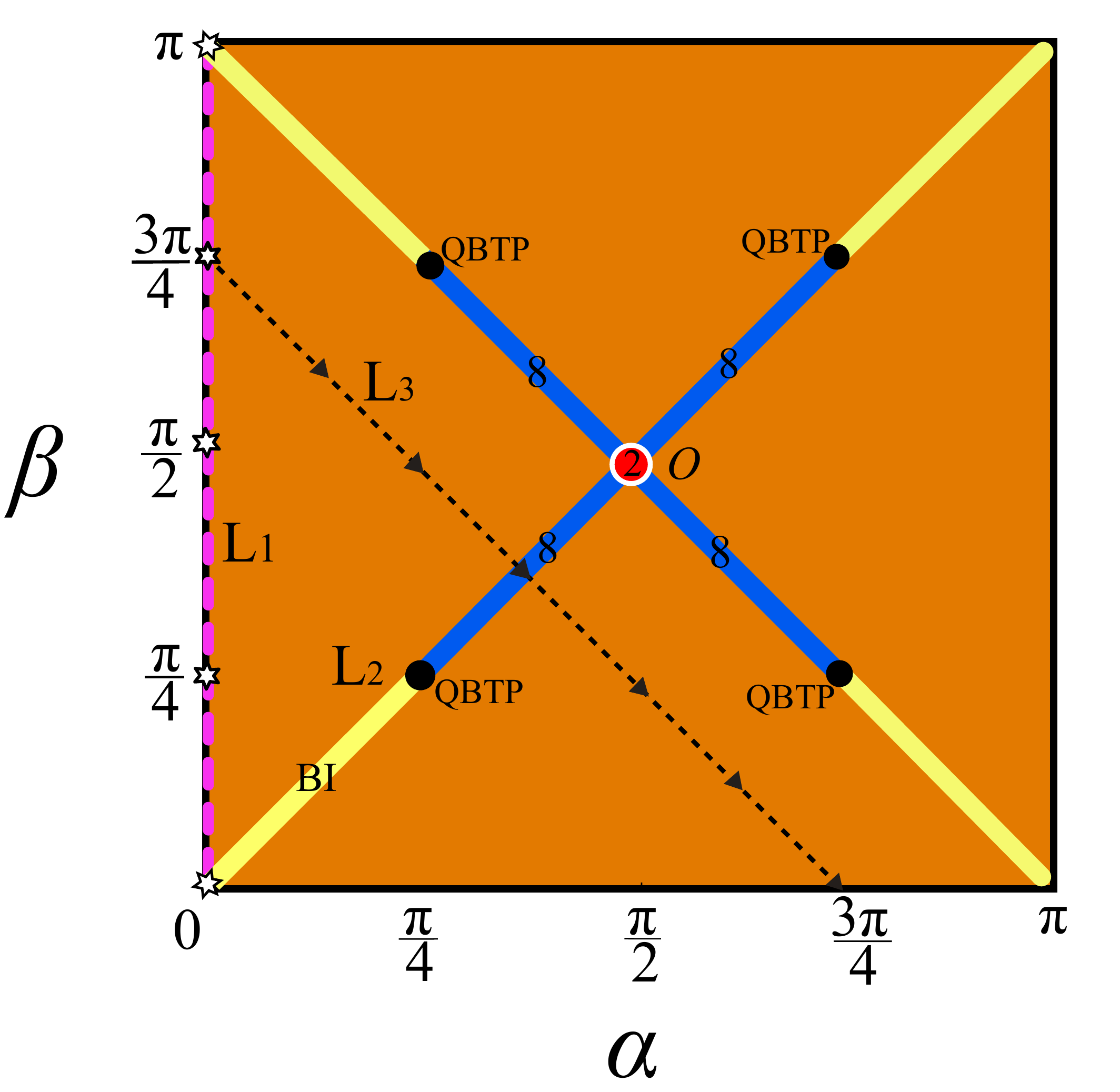}
\caption{(color online) The parameter space of the lattice model with broken inversion symmetry gauge fields in Fig.\ref{lattice} (c) for 1/4 filling. The yellow (light) thick lines indicate the trivial band insulator (BI) phases. The cyan dashed line $\mathrm{(L_1)}$, thick yellow-blue line $\mathrm{(L_2)}$ and black dashed line $\mathrm{(L_3)}$ are paths we consider in Fig.\ref{gauge1}, Fig.\ref{gauge2} and Fig.\ref{phase}, respectively. The thick blue lines correspond to phases with eight Dirac nodes, and solid circles indicate the quadratic band touching points (QBTP).  The point $O$ has 2 Dirac points in the Brillouin zone.} \label{phasediagram}
\end{center}
\end{figure}

The two special cases of broken inversion symmetry described in Fig.\ref{gauge1} and Fig.\ref{gauge2} for the flux pattern in Fig.\ref{lattice}(c) can be understood further.  Motivated by the presence of two Dirac nodes in Fig. \ref{gauge1} and eight Dirac nodes in Fig. \ref{gauge2} (see also Fig.\ref{phasediagram}), it would be interesting to look for a path in the parameter space ($\alpha,\beta$) which connect those phases (indexed by the number of Dirac nodes) through a phase transition. We choose this path as $\beta=-\alpha+3\pi/4$, and look at the deformation of the Fermi surface as the parameters are changed.  This path is shown in Fig.\ref{phasediagram} as $\mathrm{L_3}$, and for different values of gauge fields (triangles on $\mathrm{L_3}$) the Fermi surface is shown in Fig. \ref{phase}. At $\alpha=0$ the situation is already shown in Fig.\ref{gauge1}(g) and (h), where a Dirac like structure appears at the Fermi surface (for 1/4 filling).  At 1/4 filling, the Fermi level is illustrated by a red dashed line.  As the gauge fields change, bands will cross the Fermi level filling and emptying fermi pockets.  At very small values of $\alpha$ a finite Fermi surface (of holes) around the Dirac point appears, and becomes larger as the gauge fields is further increased.  We see that a Fermi pocket (of electrons) is also formed in the Brillouin zone (between $\mathrm{M'}$ and $\Gamma$) . The structure of the Fermi surface will be deformed through a band inversion close to the Fermi level. This deformation leads to shrinking of the extended Fermi surface (around the Dirac point) to point like structures as shown in Fig.\ref{phase}(c). It is interesting to note that through this evolution eight Dirac nodes are created at the Fermi level corresponding to Fig.\ref{gauge2}(g) and (h). Thus, we see how a phase transition occurs between two phases each characterizing by different numbers of Dirac nodes. As the gauge field $\alpha$ increases further, the Fermi surfaces at different parts of the Brillouin zone are created and eventually at $\alpha=3\pi/4$, Fig. \ref{phase}(f), develops two Dirac points  as seen in Fig. \ref{gauge1}(g) and (h).  (The points $\alpha=3\pi/4,\beta=0$ and $\alpha=0,\beta=3\pi/4$ have similar properties.)

To summarize this section, our main results are the following.  We have studied the Hamiltonian \eqref{H0} with gauge fields applied to the nearest neighbor links, as shown in Eq. \eqref{gaugeH}.  Thus, in this section we have taken $t=1$ in Eq.\eqref{H0} and then applied the gauge fields \eqref{unitary} to the links using the patterns illustrated in Fig.\ref{lattice} (b) and (c).  We found that when the gauge fields preserve the lattice inversion symmetry (they preserve time-reversal symmetry by construction) and a gap opens, the resulting state is a $Z_2$ TBI.  Examples of TBI are shown in Fig. \ref{edgegauge} (a) (which actually has inversion symmetry broken), (b) and (c). The TBI obtained in this section can be compared with the TBI obtained in Sec.\ref{model} where the gauge fields \eqref{unitary} were absent and instead a second neighbor spin-orbit coupling \eqref{H0} was used to open a gap and drive the state into the TBI phase.  Thus, if complex nearest-neighbor hopping parameters (that preserve time-reversal symmetry) are present a $Z_2$ TBI can be stabilized.  We speculate that it is possible to adiabatically continue a number of these states into each other (provided they have the same number of Dirac nodes on the edge) via a deformation of the Hamiltonian similar in spirit to that used in Ref.[\onlinecite{Ruegg:prb10}].  

In the final part of this section we turned our attention to the flux patterns in Fig.\ref{lattice} (c) that break inversion symmetry.  Here we found gapped states [some topological as shown in Fig. \ref{edgegauge}(a)] and gapless states depending on their strength.  In the case of broken inversion symmetry we also found phase transitions in which the number of Dirac points in the Brillouin zone could change from 2 to 8, and we described these transitions in detail. We summarize them in the Fig. \ref{phasediagram}. Different parts of the parameter space can be distinguished by looking at the deformation of the Fermi surface. The border and center of the parameter space $O$ represents the phase of the system with 2 Dirac nodes, and the thick blue lines represent the case with 8 Dirac nodes. However, most region of parameter space is characterized by having a finite fermi surface. In fact, each band insulator phase will end up with a QBTP, and these points (solid circles) evolve into eight Dirac nodes along the blue lines.  

\section{Entanglement analysis of the topological insulating phases}
\label{sec:entanglement}

In this section, we further study the TBI phases via the entanglement spectrum.\cite{Li:prl09,Thomale:prl10,Flammia:prl09,Pollmann:prb10}  The application of this method to topological band insulators has been explored earlier,\cite{Fidkowski:prl10,Turner:prb10} and we closely follow those works here.  A key result of this paper is that a tight-binding model on the square-octagon lattice with only a single orbital per site supports TBI phases at 1/4 and 3/4 filling if second neighbor spin-orbit hopping of the Kane-Mele type is included (see Sec. \ref{model}), or even if only first neighbor hopping is allowed provided non-Abelian gauge fields are added to the links of the lattice (see Sec. \ref{non-Abelian}).  We also noted that since the TBI obtained in each case has a single Dirac node on each edge, one expects them to be adiabatically connected via a deformation of the Hamiltonian similar to that used in Ref. [\onlinecite{Ruegg:prb10}]. 

In this section we return to the topological insulator phases discussed in Sec. \ref{model} (similar results would be obtained if we considered the models used in Sec. \ref{non-Abelian}) and
further characterize them by looking at the entanglement
spectrum,\cite{Li:prl09,Thomale:prl10,Flammia:prl09,Pollmann:prb10} i.e. the spectrum of the reduced density matrix for a portion of the system. Let $\rho$ be the pure density matrix of
the whole system (obtained from its wavefunction).  Given $\rho$, we divide the system into two spatially distinct regions, $A$ and $B$. The entanglement entropy is defined as
$S_A=-\mathrm{Tr}[\rho_A\ln\rho_A]$, where $\rho_A$ is the reduced density matrix defined by tracing out the degrees of freedom of part $B$, that is, $\rho_A \equiv \mathrm{Tr}_B[\rho]$. 

The reduced density matrix $\rho_A$ is a versatile tool in both quantum information theory and condensed matter physics.\cite{Eisert:rmp10,Amico:rmp08} The entanglement entropy $S_A$ obtained from it is a useful measure of the quantum correlations between two parts of the system.  Importantly, it exhibits different behavior on and off criticality.\cite{Vidal:prl03,Refael:prl04}  While it develops a logarithmic scaling in critical systems, it gets saturated for gapped systems and exhibits the so-called area law behavior there.\cite{Eisert:rmp10,cardy} Since the ground states of topologically ordered states are highly entangled, the entanglement entropy also provides some information about the topological order, namely the quantum dimension.\cite{Kitaev:prl06,Levin:prl06}

However, the spectrum of the reduced density matrix, from which the
entanglement entropy is extracted, contains richer information
about topological phases.\cite{Li:prl09,Thomale:prl10,Flammia:prl09,Pollmann:prb10} In this section we study the effect of various symmetry breaking perturbations on the entanglement spectrum of topological insulators.\cite{Fidkowski:prl10,Turner:prb10}  This approach is complementary to the study of the stability of topological band insulators discussed in Refs. [\onlinecite{Kane2},\onlinecite{Ruegg:prb10},\onlinecite{Guo:prb09}].  

 A key difference between a topological insulator and a trivial insulator is that the former has ``protected" gapless edge modes;\cite{Xu:prb06,Wu:prl06} this feature will show up in the entanglement spectrum.\cite{Fidkowski:prl10,Turner:prb10}  Since the Hamiltonian in Eq.(\ref{squareH}) is noninteracting, the reduced density matrix of any part of the system
can be fully described in terms of correlation functions as
follows,\cite{peschel1,peschel2,Cheong:prb04}
\bea \label{entaglementH}
\rho_A=\frac{1}{\mathcal{Z}}e^{-H_e},~~~~H_e=\sum_{i,j\in
A}h_{ij}c^{\dag}_ic_j, 
\eea 
where the matrix $h$ is related to the
correlation matrix $G$ with elements $G_{ij}=\langle
c^{\dag}_ic_j\rangle$ through the relation $h=-\log G+\log(\mathbf{1}-G)$.
The partition function is determined by conditioning that
$\textrm{Tr}(\rho_A)=1$ giving $\mathcal{Z}=1/\det(\mathbf{1}-G)$.
The entanglement spectrum is then given by the set of eigenvalues of the matrix $h$. Thus, both $h$ and $G$ can be diagonalized in the same basis, and their eigenvalues are monotonically related to one another as,\cite{Turner:prb10} 
\bea 
\label{monotonic}
\frac{1}{2}-\mathrm{g}_l=\frac{1}{2}\tanh(\frac{\varepsilon_l}{2}),
\eea
where $g_l$ and $\varepsilon_l$ are eigenvalues of correlation
matrix $G$ and matrix $h$, respectively. With this identification
for the entanglement spectrum, we need only work with the spectrum of the correlation
matrix. 

To calculate the entanglement spectrum, we consider a cut on the system that preserves the translational symmetry along the cut. In this case, the momentum along the cut is a 
good quantum number and we can use it for labeling the spectrum.  Thus, the single
particle entanglement eigenvalues $\varepsilon_l$ can be labeled by the linear momentum $k$ along the cut, $\varepsilon_{l}(k)$.  In the actual calculation, we consider two
parallel cuts separated by many unit cells, and take the
spectrum along one of the cuts. Note that this cut is not a physical
cut, but only a cut separating regions where degrees of freedom are traced over to obtain the reduced density matrix $\rho_A$.  In fact, as shown in Refs.[\onlinecite{Fidkowski:prl10},\onlinecite{Turner:prb10}] the entanglement spectrum can be reconstructed from a spectrally flattened
Hamiltonian. 

\begin{figure}
\begin{center}
\includegraphics[width=8cm]{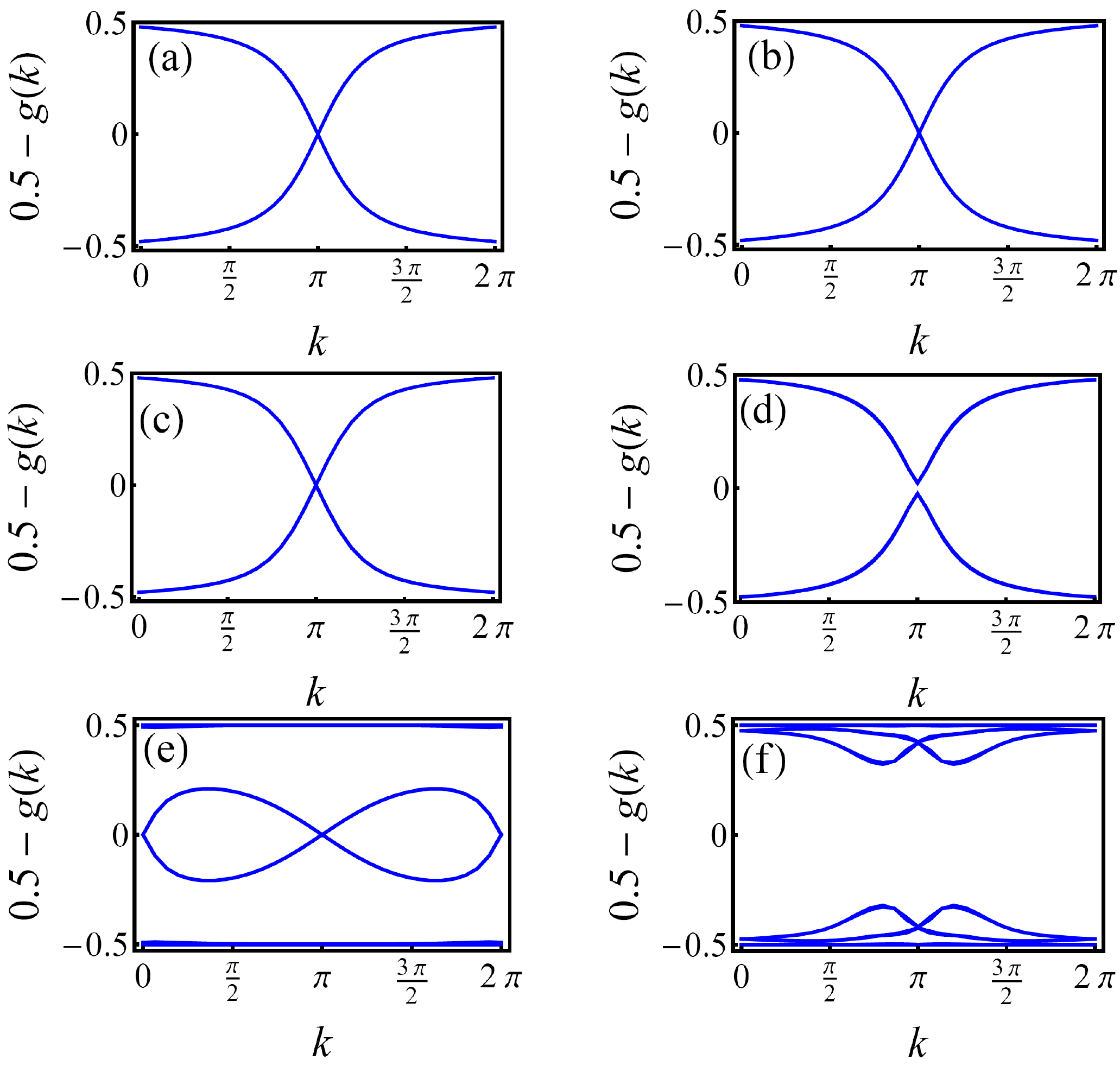}
\caption{(color online) Illustration of the entanglement spectrum for
several different cases of Eq.\eqref{squareH} with $t$ and $\lambda$ placing it in the $Z_2$ topological phase: (a) both time reversed and inversion are preserved,
(b) time reversal symmetry is broken by applying a Zeeman field
while inversion is preserved, (c) time reversal is preserved and
inversion is broken by Rashba coupling, (d) both time reversal and inversion symmetries are
broken, (e) trivial insulating phase subjected to inversion symmetry presevering sublattice potentials, and (f) trivial insulating phase with Rashba spin-orbit coupling.} \label{entanglement}
\end{center}
\end{figure}

To study the stability  via the entanglement spectrum of the $Z_2$ TBI found in this paper, we consider the model Hamiltonian in Eq.(\ref{squareH}) with the following perturbing terms: 
\bea 
\label{potential}
H_\delta=\sum_{i\sigma}\delta_ic^{\dagger}_{i\sigma}c_{i\sigma},
 \eea 
 and 
 \bea
\label{rashba}
H_\Lambda=i\Lambda\sum_{<i\alpha,j\beta>}c^{\dagger}_{i\alpha}(\overrightarrow{\sigma}_{\alpha\beta}\times\hat{d}_{ij})_{z}c_{j\beta}+\mathrm{H.c.},
\eea
where the $H_\delta$ and $H_\Lambda$ describe the charge density
modulation and Rashba spin-orbit coupling. The $\delta_i=\pm\delta$ can be chosen
to either preserve or break the inversion symmetry of the
lattice, while the Rashba coupling necessarily breaks the inversion symmetry.
We consider both cases to study the stability of the
entanglement edge modes. As it turns out in the honeycomb lattice
model\cite{Kane1} or its variant,\cite{Ruegg:prb10} these perturbations
suppress the topological insulator phase at critical values in the present square-octagon lattice as well. 

First, we study the entanglement modes of the bare model without above
perturbations. Fig. \ref{entanglement}(a) depicts how the edge modes
of the topological insulator phase is reflected in the entanglement
spectrum: the spectrum is gapless, just like the physical edge spectrum.  
The gapless physical edge modes are protected by time reversal
symmetry. However, the gapless modes in the entanglement spectrum are protected even if the time reversal symmetry is broken.\cite{Turner:prb10} To check this in the current model, we perturb
Hamiltonian by a Zeeman term as
\bea
H_z=h^z\sum_{i\sigma}c^{\dagger}_{i\sigma}\sigma_{\sigma\sigma}^{z}c_{i\sigma},
\eea
which explicitly breaks the time reversal symmetry and gaps out the physical edge
modes. However, as shown in Fig. \ref{entanglement}(b) the gapless
entanglement edge modes remain intact. This implies that the entanglement
modes enjoy a higher degree of robustness than the physical edge modes. In fact, the
entanglement edge modes are protected by inversion symmetry:\cite{Turner:prb10,Pollmann:prb10} For inversion symmetric systems, the inversion along the cut maps the right-hand part to the left-hand side of the cut. This invokes a kind of particle-hole symmetry,\cite{Turner:prb10} and results in single particle entanglement energies 
having the property $\varepsilon_{\bar{l}}(-k)=-\varepsilon_{l}(k)$. This 
relation clearly reveals that the edge modes of the entanglement
spectrum at time reversal invariant momenta are degenerate with zero energy.

However, the gapless nature of the entanglement edge spectrum can survive even if the
inversion is broken, say by Rashba term in Eq.\eqref{rashba}, provided the time reversal symmetry is not also broken, as can be seen in Fig. \ref{entanglement}(c). Indeed, gapless physical edge
modes lead to the degeneracy of the eigenvalues of the reduced density matrix\cite{Fidkowski:prl10,Turner:prb10} and thus to the gapless
entanglement edge modes.  This follows from the argument that the $2^M$ eigenvalues
of the reduced density matrix Eq.(\ref{entaglementH}) can be written
as $\prod^{M}_{l=1}[1+s_l(e^{-\varepsilon_l}-1)]$ with $s_l=0,1$.
Trivially, a zero entanglement spectrum eigenvalue makes the density matrix
spectrum degenerate. On the other hand, $\varepsilon_l$ can be
related to the edge modes of a spectrally flattened Hamiltonian
respecting the ground state and phase of the original Hamiltonian.\cite{Fidkowski:prl10,Turner:prb10}
In such a transformation, zero energy modes, which are often
confined to the boundary, are reflected in the degeneracies of the
eigenvalues of the reduced density matrix. 

Finally, breaking both time reversal and inversion symmetry gaps out the edge modes of the entanglement spectrum as shown in Fig. \ref{entanglement}(d). We see that the number of gapless edge modes in the entanglement spectrum is also consistent with the 
$Z_2$  characterization of the topological insulators.\cite{Turner:prb10} To check the consistency with non-topological insulating phases, we consider two sets of parameters corresponding to the trivial insulating phases. Fig.\ref{entanglement}(e) and  Fig.\ref{entanglement}(f) illustrate the entanglement modes for trivial insulating phase resulting from the symmetric sublattice potential $\delta=1.2$ and Rashba coupling $\Lambda=0.6$, respectively (with $t=1$) in Eq.\eqref{squareH}. Both cases show an even number of edge modes consistent with their trivial $Z_2$ values. 

\section{Summary and Conclusions}
\label{conclusions}

In this paper we introduced a single-orbital tight-binding model defined on the square-octagon lattice. First we studied a Kane-Mele type limit with spin-orbit coupling for second neighbor hopping, and showed that the model could support topological insulating phases depending on filling factors and the parameters of the model. In particular, we found that upon the changing of the spin-orbit coupling, the topological insulating phase turns into a semimetallic phase that could potentially have counterpropagating edge modes, which is qualitatively similar to bulk Sb.\cite{Fu:prb07} Then we considered a specific modification of hopping terms in the Hamiltonian by coupling them to nontrivial gauge fields on nearest neighbor links (with no second-neighbor hopping). We showed such gauge fields can give rise to $Z_2$ topological insulating phases. We found in all cases with inversion symmetry and a finite gap, the model is in a topological insulating phase, and in some cases with inversion symmetry broken.  Changing the gauge fields can also lead to topological phase transitions between different metallic phases. These metallic phases are characterized by the evolution of the Fermi surface and the number of Dirac nodes in the Brillouin zone. We also examined the stability of the topological insulating phases by using the entanglement spectrum. While the time reversal breaking perturbations gap out the physical edge modes, as long as the inversion symmetry is preserved, the entanglement edge modes remains gapless. 

By providing a new example of a lattice that supports $Z_2$ topological band insulators (realized via both second-neighbor spin-orbit coupling and nearest-neighbor non-Abelian gauge fields) in a simple s-band model, we have expanded number of known systems where these topological phases exist.  Our two approaches to achieving a topological state imply that both solid-state and cold atom systems can likely be found where the physics discussed here is relevant. 

\acknowledgements
We thank Victor Chua, Andreas R\"uegg, Ari Turner, Jun Wen, and Ashvin Vishwanath for enlightening discussions. We gratefully acknowledge financial support from ARO Grant W911NF-09-1-0527.


%

\end{document}